\documentclass[prl, reprint, floatfix]{revtex4-1}

\usepackage{amsfonts, amsmath, amssymb, bm}
\usepackage{hyperref}
\usepackage{graphicx}
\usepackage{caption, subcaption}
\usepackage{MnSymbol}
\usepackage[strict]{changepage}
\usepackage[makeroom]{cancel}
\usepackage{color}

\newcommand{\mse}{\mathsf{mse}}

\begin{document}

\title{Online compressed sensing}
\date{\today}
\author{Paulo V. Rossi}
\email{pvcrossi@if.usp.br}
\affiliation{Dep.~de F\'{i}sica Geral, Instituto de F\'{i}sica, University of S\~{a}o Paulo, \\
S\~{a}o Paulo-SP 05314-090, Brazil}
\author{Yoshiyuki Kabashima}
\email{kaba@dis.titech.ac.jp}
\affiliation{
Department of Computational Intelligence and Systems Science, \\
Tokyo Institute of Technology,
Yokohama 226-8502, Japan}
\author{Jun-ichi Inoue}
\affiliation{Graduate School of Information Science and Technology, \\
Hokkaido University, N14-W9, Kita-ku, Sapporo 060-0814, Japan}

\begin{abstract}
In this paper, we explore the possibilities and limitations of recovering sparse signals in an online fashion.
Employing a mean field approximation to the Bayes recursion formula yields an online signal recovery algorithm that can be performed with a computational cost that is linearly proportional to the signal length per update.
Analysis of the resulting algorithm indicates that the online algorithm asymptotically saturates the optimal performance limit achieved by the offline method in the presence of Gaussian measurement noise, while differences in the allowable computational costs may result in fundamental gaps of the achievable performance in the absence of noise.
\end{abstract}
\maketitle

%
%
%
%

Continuous innovations in measurement technologies have enabled the collection of high-dimensional data of various objects.
This trend has created new research areas such as bioinformatics that apply knowledge and techniques from information science to the natural sciences in order to efficiently extract information  from data.
The importance of techniques for efficient information extraction has been growing more than ever in various fields of science and engineering \cite{Cyber05,fourth_paradigm,E-science}.

Compressed sensing (CS), which is a framework for signal processing that is currently under development, is a successful resource that has produced many such techniques \cite{Donoho06compressedsensing,Candes08anintroduction}.
In general, CS aims to realize high-performance signal processing by exploiting the prior knowledge of objective signals, particularly their {\em sparsity}. 
That is, CS utilizes the fact that real-world signals can typically be represented by a small combination of elemental components.
In a standard scenario, CS utilizes this property to enable the recovery of various signals from much fewer samples of linear measurements than required by the Nyquist--Shannon theorem \cite{Donoho06,DonohoTanner09,CandesTao05,CandesTao06,KWT09,DMM09,Gangli10,krzakala,RFG12}.

Besides the standard scenario, the concept of CS is now spreading in various directions.
For instance, in a remote sensing situation where a sensor transmits data to a data center placed at a far distance, the amount of data to be sent through a narrowband communication channel may be the biggest hindrance to efficient signal processing.
As a practical solution to such a difficulty, 1-bit CS is a proposed scheme for recovering sparse signals by using
only the sign information of the measurement results \cite{BB_CISS08,yingying,yingying2}.
Another variation can be utilized when multiple signals are observed in a distributed manner.
In such cases, the signal recovery performance can also be enhanced by exploiting information on correlations among the signals.
This is referred to as distributed CS \cite{Duarte05distributedcompressed,Baron09,shiraki15}.

In this letter, we explore the possibilities and limitations of another variation of CS, which we call {\em online CS}.
In this scheme, to minimize the computation and memory costs as much as possible, measured data are used for signal recovery only once and discarded after that.
This approach towards information processing is a promising technique for when devices of low computational capability are used for signal recovery; such situations can arise in sensor networks, multi-agent systems, 
micro-devices,
and so on.
It is also advantageous when the signal source is time-variant.

Historically, online information processing was actively investigated by the physics community more than two decades ago in the context of learning by neural networks \cite{kinouchi92,biehl94,kabashima94,saad95,opper96,vandenbroeck96,vicente98,Saad98}.
However, the utility of sparsity was not fully recognized at that time, so the potential capability of online CS remains an open question.
To clarify this issue, we focused on the performance when the Bayesian inference is considered in an online manner, which is guaranteed to yield the optimal performance when the signal recovery is carried out in an offline (batch) manner.

%
%
%
%
\paragraph{Problem setup.}
As a general scenario, we 
consider
a situation where an $N$-dimensional signal $\bm{x}^0=(x_{i}^0) \in \mathbb{R}^N$ is sequentially measured by taking the inner product 
$u^{0,t}=\bm{\Phi^t}\cdot \bm{x}^0$
for a random measurement vector $\bm{\Phi^t}=(\Phi_{i}^t)\in \mathbb{R}^N$.
Here, we assumed that each component of 
$\bm{x^0}$
is independently generated from an identical sparse distribution $\phi(x)=(1-\rho)\delta(x)+\rho f(x)$ and that each component of $\bm{\Phi}^t$ independently follows a distribution of zero mean and variance $N^{-1}$, where $0<\rho<1$, $i=1,2,\ldots, N$ and $f(x)$ is a density function that does not have a finite mass at the origin.
The index $t=1,2,\ldots$ counts the number of measurements.
For each measurement, the output $y^t$, which may be a continuous or discrete variable, is sampled from 
a
conditional distribution 
$P(y^t|u^{0,t})$.
Here, our goal is to accurately recover $\bm{x}^0$ based on the knowledge of $D^t=\{(\bm{\Phi}^1,y^1),(\bm{\Phi}^2,y^2),\ldots,(\bm{\Phi}^t,y^t)\}$ and the functional forms of $\phi(x)$ and $P(y|u)$ while minimizing the necessary computational cost.

\paragraph{Bayesian signal recovery and online algorithm.}
Let $\hat{\bm{x}}(D^t)$ denote the estimate of $\bm{x}^0$ by an arbitrary recovery scheme.
The standard measure of the accuracy of $\hat{\bm{x}}(D^t)$ is the mean square error (MSE) $\mse = N^{-1}
\left [ ||\hat{\bm{x}}(D^t) -\bm{x}^0 ||^2 \right ]_{D^t,\bm{x}^0}$, where $\left [ \cdots \right ]_X$ generally indicates the average operation with respect to $X$.
Fortunately, under the current setup, Bayes' theorem, which is given by
\begin{eqnarray}
P(\bm{x}|D^t)=\frac{\prod_{\mu=1}^t P(y^\mu|u^\mu) \prod_{i=1}^N \phi(x_i)}
{\int d\bm{x} \prod_{\mu =1}^t P(y^\mu |u^\mu) \prod_{i=1}^N \phi(x_i)},
\label{Bayes}
\end{eqnarray}
guarantees that $\mse$ is minimized by the minimum mean square error estimator
$\hat{\bm{x}}^{\rm mmse}(D^t) =\int d \bm{x} \bm{x} P(\bm{x}|D^t)$.
However, evaluating $\hat{\bm{x}}^{\rm mmse}(D^t)$ exactly is, unfortunately, computationally difficult.

To practically resolve this difficulty, we 
introduce
the following two devices:
\begin{itemize}
\item {\bf Online update:} We rewrite (\ref{Bayes}) in the form of a recursion formula:
\begin{eqnarray}
P(\bm{x}|D^{t+1})=\frac{P(y^{t+1}|u^{t+1}) P(\bm{x}|D^t)}
{\int d \bm{x} P(y^{t+1}|u^{t+1}) P(\bm{x}|D^t)}.
\label{recursion}
\end{eqnarray}
\item {\bf Mean field approximation:} To make the necessary computation tractable, we approximate $P(\bm{x}|D^t)$ by 
 a  factorized distribution of the exponential family \cite{Amari07} as
\begin{eqnarray}
P(\bm{x}|D^t) \simeq \prod_{i=1}^N \left (
\frac{e^{-a_i^tx_i^2/2 + h_i^t x_i} \phi(x_i)}
{Z(a_i^t,h_i^t)} \right)
\label{natural_param}
\end{eqnarray}
while utilizing the set of natural parameters $\{(a_i^t,h_i^t)\}$, where $Z(a_i^t,h_i^t)=\int dx_i e^{-a_i^t x_i^2/2 + h_i^t x_i} \phi(x_i) $.
\end{itemize}

Introducing online computation to the Bayesian inference based on conversion from (\ref{Bayes}) to (\ref{recursion}) has also been proposed in earlier studies on the learning of neural networks \cite{opper98,winther98}.
On the other hand, the parameterization of (\ref{natural_param}) for the approximate tractable distribution
may not have been popular for the online learning of neural networks.

When the prior distribution of $\bm{x}$ is smooth, which is typically the case in neural network models, the posterior distribution 
is expected to asymptotically approach
a Gaussian distribution.
This means that, at least in the asymptotic region of $\alpha =t/N \gg 1$, the posterior distribution can be closely approximated as $\displaystyle P(\bm{x}|D^t) \propto \exp \left (-(\bm{x}-\bm{m}^t)^{\rm T} (C^t)^{-1} (\bm{x}-\bm{m}^t)/2 \right)$ by employing the mean $\bm{m}^t=\int d\bm{x} \bm{x} P(\bm{x}|D^t)$ and covariance $C^t= \int d\bm{x} (\bm{x}\bm{x}^{\rm T})P(\bm{x}|D^t) - \bm{m}\bm{m}^{\rm t}$ as parameters, where $\rm T$ denotes the matrix transpose.
Supposing this property, earlier studies derived update rules directly for $\bm{m}^t$ and $C^t$.
However, in the current case, the strong singularity of the prior $\phi(x)$, which originates from the component of $\delta(x)$, prevents $P(\bm{x}|D^t)$ from converging to a Gaussian distribution, even for $\alpha \gg 1$.
To overcome this inconvenience, we derived update rules for $\{(a_i^t, h_i^t)\}$ based on the expression of (\ref{natural_param}) and computed the means and variances as $m_i^t=(\partial/\partial h_i^t)\ln Z(a_i^{t},h_i^{t})$ and $v_i^t=(\partial^2/(\partial h_i^t)^2)\ln Z(a_i^{t},h_i^{t})$, respectively.

Let $\{(a_i^t, h_i^t)\}$ be given; therefore, $\{(v_i^t,m_i^t)\}$ is also provided.
The update rule of $(a_i^t,h_i^t)\to (a_i^{t+1},h_i^{t+1})$ is derived by inserting the expression of (\ref{natural_param}) to the right-hand side of (\ref{recursion}) and integrating the resultant expression with respect to $\bm{x}$ except for $x_i$.
In the integration, we approximate $u_{\backslash i}^{t+1} =\sum_{j \ne i}\Phi_j^{t+1} x_j$ as a Gaussian random variable whose mean and variance are $\Delta_{\backslash i}^{t+1} =\sum_{j \ne i} \Phi_j^{t+1} m_j^{t}$ and $\chi_{\backslash i}^{t+1}= \sum_{j \ne i} (\Phi_j^{t+1})^2 v_j^{t} \simeq N^{-1} \sum_{j \ne i} v_j^{t}$, respectively.
This is supported by the assumption for the distribution of the measurement vectors $\bm{\Phi}^t$.
By employing this Gaussian approximation to evaluate the integral and expanding
the 
resultant expression up to
the second order in $\Phi_i^{t+1} x_i$, we can obtain the online signal recovery algorithm as follows:

\begin{widetext}
\begin{equation}
\left \{
\begin{array}{l}
\displaystyle
a_i^{t+1} = a_i^t - (\Phi_i^{t+1})^2
\frac{\partial^2}{(\partial \Delta^{t+1})^{2}}
\ln \! \int \! \mathcal Dz P(y^{t+1}|
\Delta^{t+1} + \sqrt{\chi^{t+1}} \,z) , \cr
\displaystyle
h_i^{t+1} = h_i^t + \Phi_{i}^{t+1}
\frac{\partial}{\partial \Delta^{t+1}} \ln \! \int \! \mathcal Dz P(y^{t+1}|
\Delta^{t+1} + \sqrt{\chi^{t+1}} \,z) - m_i^t(\Phi_i^{t+1})^2 \frac{\partial^2}{(\partial \Delta^{t+1})^{2}}
\ln \! \int \! \mathcal Dz P(y^{t+1}|
\Delta^{t+1} + \sqrt{\chi^{t+1}} \,z) ,
\end{array}
\right .
\label{eq:general-micro_dynamics}
\end{equation}
\end{widetext}
where $\Delta^t = \sum_{i=1}^N\Phi_i^{t}m_i^{t-1}$,
$\chi^t = \sum_{i=1}^N (\Phi_i^{t})^2v_i^{t-1}$ and
$\mathcal Dz=dz\exp(-z^2/2)/\sqrt{2\pi}$ represents the Gaussian measure.
Note that the necessary cost of computation for performing (\ref{eq:general-micro_dynamics}) is $O(N)$ per update.
This means that the {\em total} computational cost for the recovery when using $t=\alpha N$ measurements is $O(N^2)$, which is comparable to the cost {\em per update} of existing fast offline signal recovery algorithms \cite{DMM09,krzakala,rangan10}.

\paragraph{Macroscopic analysis.}
Because $\bm{\Phi}^t$ and $y^t$ are random variables, (\ref{eq:general-micro_dynamics}) constitutes a pair of stochastic difference equations.
However, because $\Phi_i^t\sim O(N^{-1/2})$, the difference with each update becomes 
infinitesimally small
as $N$ grows.
This property makes it possible to reduce (\ref{eq:general-micro_dynamics}) to a set of ordinary differential 
equations 
\begin{widetext}
\begin{equation}
\left \{
\begin{array}{l}
\displaystyle
\frac{d \hat{Q}}{d \alpha} =- \mathop{\rm Tr}_y
\int \mathcal D v \int \mathcal D u P\Big (y \Big | \frac{m}{\sqrt{q}} \, v + \sqrt{Q_0-\frac{m^2}{q}} \, u \Big)
\frac{\partial^2}{(\partial \sqrt{q} v)^2} \ln
 \int \mathcal Ds P(y| \sqrt{q} \, v + \sqrt{Q-q} \, u), \cr
 \displaystyle
 \frac{d \hat{q}}{d \alpha} =\mathop{\rm Tr}_y
\int \mathcal D v \int \mathcal Du P\Big (y \Big | \frac{m}{\sqrt{q}} \, v + \sqrt{Q_0-\frac{m^2}{q}} \, u \Big)
\left (\frac{\partial}{\partial \sqrt{q} t} \ln
 \int \mathcal Ds P(y| \sqrt{q}\, v + \sqrt{Q-q} \, u) \right)^2, \cr
 \displaystyle
 \frac{d \hat{m}}{d \alpha} = \mathop{\rm Tr}_y
 \int \mathcal D v
\left (\frac{\partial}{\partial (m v /\sqrt{q})}
\int \mathcal Du P\Big (y \Big | \frac{m}{\sqrt{q}} \,v + \sqrt{Q_0-\frac{m^2}{q}} \, u \Big) \right)
\left (\frac{\partial}{\partial \sqrt{q} v} \ln
 \int \mathcal Du P(y| \sqrt{q} \, v + \sqrt{Q-q} \, u) \right),
 \end{array}
 \right .
 \label{eq:macrodynamics}
 \end{equation}
\end{widetext}
in the limit of $N, t \to \infty$ but keeping $\alpha=t/N$ finite,
 where $Q_0=\int dx\phi(x)x^2$, $\mathop{\rm Tr}_y$
denotes the integration or summation with respect to $y$, and $q$, $m$, and $Q$ are evaluated as $q=\int dx^0 \phi(x^0) \mathcal D z \left \langle x \right \rangle^2$, $m=\int dx^0 \phi(x^0) \mathcal D z x^0 \left \langle x \right \rangle$, and $Q=q+\int dx^0 \phi(x^0) \mathcal D z \partial \left \langle x \right \rangle/\partial(\sqrt{\hat{q}}z)$ using $\left \langle x \right \rangle =(\partial/\partial(\sqrt{\hat{q}}z))
\ln Z(\hat{Q},\sqrt{\hat{q}}z+\hat{m}x^0)$.

Two issues are of note here.
First, replacing $(d\hat{Q}/d\alpha, d\hat{q}/d \alpha, d\hat{m}/d \alpha)$ with $(\hat{Q}/\alpha, \hat{q}/\alpha, \hat{m}/\alpha)$ in (\ref{eq:macrodynamics}) yields the exact equation of state for the Bayesian {\em offline} signal recovery, which is derived by the replica or cavity method \cite{krzakala,yingying2}.
This implies that the differences in the macroscopic descriptions---i.e., the use of differential instead of algebraic equations---characterize the fundamental limitations on the achievable performance of 
the online method 
(\ref{eq:general-micro_dynamics})
compared to 
the offline method.
Second, similar to the Bayes optimal case for the offline recovery, the equation of state (\ref{eq:macrodynamics})
allows a solution with $\hat{Q}=\hat{q}=\hat{m}$, $Q=Q_0$, and $q=m$.
Focusing on the solution of this type simplifies (\ref{eq:macrodynamics}) to
\begin{eqnarray}
&&\frac{d \hat{q}}{d \alpha}
=\mathop{\rm Tr}_y
\int \mathcal D v \int \mathcal Du P (y | \sqrt{q} \,
v + \sqrt{Q_0-q} \, u) \cr
&& \ \ \ \ \times
\left (\frac{\partial}{\partial \sqrt{q} v} \ln
 \int \mathcal Du P(y| \sqrt{q}\,v+ \sqrt{Q_0-q} \, u) \right)^2,
 \label{eq:Nishimori}
\end{eqnarray}
where $q=\int dx^0 \phi(x^0) \mathcal D z \left (\partial/\partial (\sqrt{\hat{q}} z) \ln Z(\hat{q},\sqrt{\hat{q}} z+\hat{q} x^0) \right)^2$.
Because the numerical computation indicate that this solution is the unique attractor of (\ref{eq:macrodynamics}), we examined the performance of the online algorithm by utilizing (\ref{eq:Nishimori}).

\paragraph{Examples.}
We tested the developed methodologies on two representative scenarios of CS.
The first is the standard CS, which is characterized by $P(y|u)=(2 \pi \sigma_n^2)^{-1/2}\exp \left (-(y-u)^2/(2 \sigma_n^2) \right)$. 
The other is 
the
1-bit CS, which 
is modeled by $P(y|u) = \int \mathcal Dz \Theta \left (yu+\sigma_n z \right)$.
Here, $y \in \{+1,-1\}$, and $\Theta(x)=1$ for $x \ge 0$ and $0$ otherwise.
For practical relevance, we considered situations where each measurement was degraded by Gaussian noise of zero mean and variance $\sigma_n^2$ for both cases.
However, by setting $\sigma_n^0 = 0$, we 
can
also evaluate the performance of a noiseless setup.
For the generative model of sparse signals, we considered the case of the Bernolli--Gaussian prior
$\phi(x)=(1-\rho)\delta(x)+\rho(2 \pi \sigma^2)^{-1/2} \exp \left (-x^2/(2 \sigma^2) \right) $,
which means that $Q_0=\rho \sigma^2$.

\begin{figure*}[t]
\centering
		\includegraphics[width=0.90\textwidth]{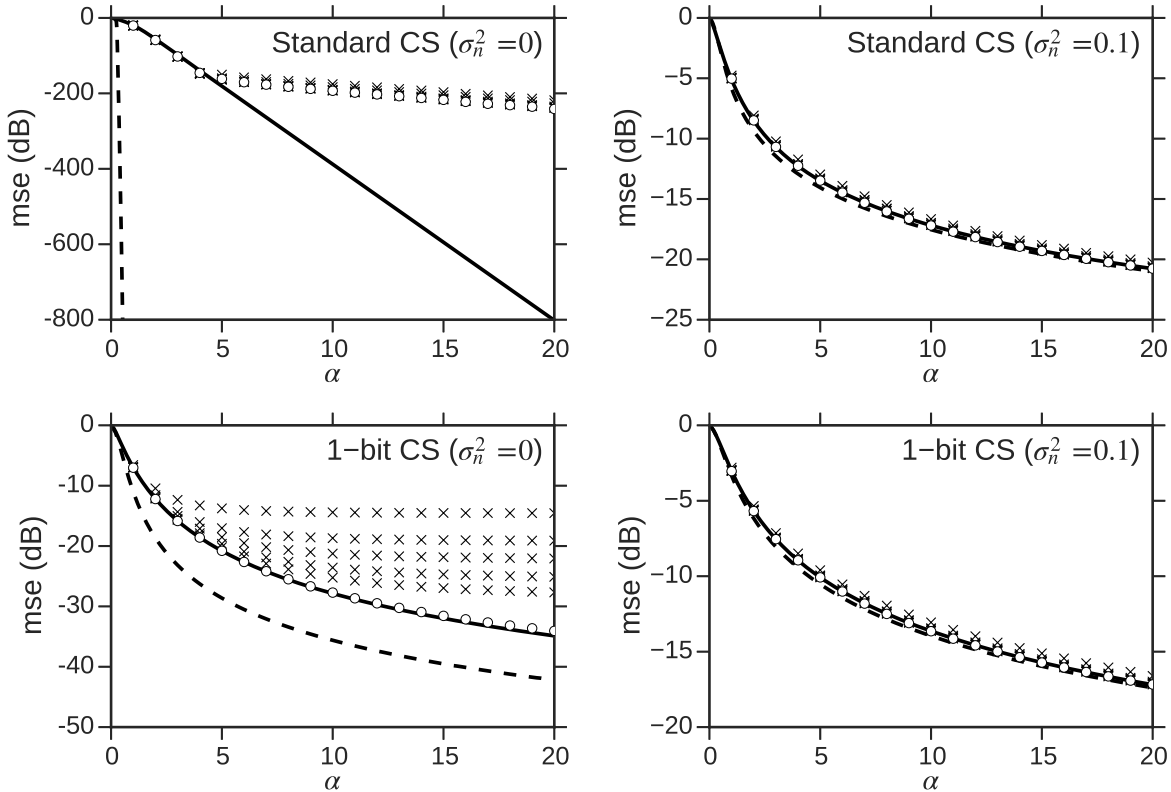}
		\caption{Comparison between $\mse$ from experimental results and theoretical predictions for $\rho = 0.1$. The crosses correspond to $N=200, 500, 1000, 2000$, and $4000$ (in descending order), and the white circles are the extrapolations of the data to $N\to\infty$ by quadratic fitting. The curves represent the theoretical performances of the online (continuous) and batch (dashed) 
reconstructions.
The disagreement between the experimental and theoretical results in the noiseless standard CS case was due to the limited numerical accuracy of the computational environment used in this study.
Also in this case, batch 
reconstruction achieves
$\mse=0$ for $\alpha$ larger than $\alpha_{\rm c}(\rho)<1$.\label{theory_vs_experiments}}
	\end{figure*}

Fig. \ref{theory_vs_experiments} compares $\mse$ from the experimental results obtained with (\ref{eq:general-micro_dynamics}) and the theoretical predictions.
The experimental results represented averages over 1000 samples, while the theoretical predictions were evaluated by solving (\ref{eq:Nishimori}) with the use of the Runge--Kutta method.
With the exception of noiseless standard CS, where numerical accuracy becomes an issue because of the 
extremely 
small values of $\mse$, the experimental data extrapolated to $N \to \infty$ exhibited excellent agreement with the theoretical predictions.
Note that the data of finite $N$ were biased monotonically to be higher for smaller $N$ and larger $\alpha$.
	
For noiseless standard CS, the 
offline reconstruction achieves
$\mse=0$ when $\alpha$ is
greater than a certain
critical ratio $0 < \alpha_{\rm c}(\rho) < 1$ \cite{krzakala}.
On the other hand, the analysis based on (\ref{eq:Nishimori}) indicated that $\mse \simeq O\left (\exp(-\alpha /\rho) \right) $ holds for large $\alpha$, which means that perfect recovery is unfortunately impossible with (\ref{eq:general-micro_dynamics}).
However, this result may still promote the use of the online algorithm for very sparse signals 
with 
$0 < \rho\ll 1$ where $\exp(-\alpha /\rho)$ becomes negligible.

For noiseless 1-bit CS, the result of \cite{yingying2} meant that $\mse \simeq (Q_0/2) \left (\frac{\rho}{K\alpha} \right)^2$ was asymptotically achieved by the offline method, where $K=0.3603\ldots$.
On the other hand, (\ref{eq:Nishimori}) yielded the asymptotic form $\mse \simeq 2 Q_0 \left (\frac{\rho}{K\alpha} \right)^2$ for $\alpha \gg 1$.
This indicates that online recovery can save computation and memory costs considerably while sacrificing $\mse$ by only a factor $4$ asymptotically.

These results may imply that there are fundamental gaps in the asymptotically achievable performance limit
depending on the allowable computational costs in the absence of noise.
However, this is not the case when Gaussian measurement noise is present, for which $P(y|u)$ becomes differentiable with respect to $u$.
This property guarantees that $\hat{q} \simeq I \alpha$ asymptotically holds for both the online and offline methods,
which yields a universal expression for the asymptotic MSE: $\mse \simeq 2\rho/\hat{q}=2 \rho/(I\alpha)$, where
\begin{eqnarray}
I =\mathop{\rm Tr}_{y} \!\! \int \!\!
\mathcal D v P\left (y \Big |\sqrt{Q_0} v\right)
\left (\frac{\partial}{\partial \sqrt{Q_0}v}\ln
P\left (y \Big |\sqrt{Q_0} v\right) \right)^2
\label{eq:FisherInformation}
\end{eqnarray}
represents the Fisher information of the measurement model $P(y|u)$ averaged over the generation of $\bm{\Phi}^t$.
This impresses the potential utility of the online algorithm and indicates that a performance similar to
that of the offline method can be asymptotically achieved with a significant reduction in computational costs.

\paragraph{Summary and discussion.}We developed an online algorithm to 
perform
Bayesian inference on the signal recovery problem of CS. The algorithm can be carried out with $O(N)$ computational and memory costs per update, which are considerably less than those of the offline algorithm.
From the algorithm, we also derived
ordinary differential equations with respect to macroscopic variables 
that were utilized for the performance analysis.
Our analysis indicated that the online algorithm can asymptotically achieve the same MSE as the offline algorithm with a significant reduction of computational costs in the presence of Gaussian measurement noise, while there may exist
certain fundamental gaps in the achievable performance depending on usable computational resources in the absence of noise.
Numerical experiments based on the standard and 1-bit scenarios supported our analysis.

Here, we assumed that correct knowledge about the prior distribution of signals and the measurement model is provided
in order to evaluate the potential ability of the online algorithm.
However, such information is not necessarily available in practical situations.
Incorporating the idea of online inference into situations lacking correct prior knowledge is an important and
challenging future task.

PVR acknowledges FAPESP for supporting his stay at the Tokyo Institute of Technology under Grant No. 2014/22258-2.
This work was partially supported by JSPS KAKENHI No. 25120013 (YK).

%

\end{document}